\documentclass[iop]{emulateapj}
\usepackage{amsmath}
\usepackage{graphicx}
\usepackage{psfrag}
\usepackage{color}
\usepackage{dcolumn}
\usepackage{bm}
\usepackage{longtable}
\usepackage[mathscr]{eucal}
\usepackage{mathrsfs}
\usepackage{xspace}

\def\msun{\rm M_{\odot}}

\def\simlt{\mathrel{\rlap{\lower 3pt\hbox{$\sim$}}\raise 2.0pt\hbox{$<$}}}
\def\simgt{\mathrel{\rlap{\lower 3pt\hbox{$\sim$}} \raise 2.0pt\hbox{$>$}}}
\def\lsim{\mathrel{\rlap{\lower 3pt\hbox{$\sim$}}\raise 2.0pt\hbox{$<$}}}
\def\gsim{\mathrel{\rlap{\lower 3pt\hbox{$\sim$}} \raise 2.0pt\hbox{$>$}}}

\def\msunpc3{\msun~{\rm {pc^{-3}}}}
\newcommand{\be}{\begin{equation}}
\newcommand{\ee}{\end{equation}}

\newcommand{\bea}{\begin{eqnarray}}
\newcommand{\eea}{\end{eqnarray}}
\newcommand{\beq}{\begin{equation}}
\newcommand{\eeq}{\end{equation}}
\newcommand{\KMS}{\rm km/s}

\begin{document}

\def\fun#1#2{\lower3.6pt\vbox{\baselineskip0pt\lineskip.9pt
  \ialign{$\mathsurround=0pt#1\hfil##\hfil$\crcr#2\crcr\sim\crcr}}}
\def\lap{\mathrel{\mathpalette\fun <}}
\def\gap{\mathrel{\mathpalette\fun >}}
\def\kms{\rm km/s}
\def\vk{V_{\rm recoil}}

\title{Modeling the Black hole Merger of QSO 3C 186}

\author{Carlos O.Lousto, Yosef Zlochower, and Manuela Campanelli}
\email{lousto@astro.rit.edu}
\affiliation{Center for Computational Relativity and Gravitation,\\
and School of Mathematical Sciences, Rochester Institute of
Technology, 85 Lomb Memorial Drive, Rochester, New York 14623}

\begin{abstract}
  Recent detailed observations of the radio-loud quasar 3C 186
  indicate 
  the possibility that a supermassive recoiling black hole is moving
  away from the host galaxy at a speed of nearly 2100km/s. If this is
  the case, we can model the mass ratio and spins of the progenitor binary black hole
using the results of numerical relativity
  simulations. We find that the black holes in the  progenitor must have
  comparable masses with a mass ratio  $q=m_1/m_2>1/4$ and the spin of
  the primary black hole must be $\alpha_2=S_2/m_2^2>0.4$. The final remnant of
  the merger is bounded by $\alpha_f>0.45$ and at least $4\%$ of the total
  mass of the binary system is radiated into gravitational waves.
  We consider four different pre-merger scenarios that further narrow
  those values.
  Assuming, for instance, a cold accretion driven merger model, we find
  that the binary had comparable masses with $q=0.70^{+0.29}_{-0.21}$
  and the normalized spins of the larger and smaller black holes were
  $\alpha_2=0.94^{+0.06}_{-0.22}$ and $\alpha_1=0.95^{+0.05}_{-0.09}$.
  We can also estimate the final recoiling black hole spin
  $\alpha_f=0.93^{+0.02}_{-0.03}$ and that the system radiated
  $9.6^{+0.8}_{-1.4}\%$ of its total mass, making the merger of those
  black holes the most energetic event ever observed.


\end{abstract}

\maketitle

\section{Introduction}\label{sec:Introduction}

A recent detailed study \cite{Chiaberge:2016eqf} of the radio-loud
quasar 3C 186, which has an active nucleus offset from the galactic
center by $1.3 \pm 0.1$ arcsec (i.e. $\sim11$ kpc) and broad line
emissions offset from the narrow line spectra by $2140\pm390$ km/s,
has concluded that the most likely explanation  is that the central
supermassive black is recoiling away from the center of the galaxy at
$\sim2000$ km/s.

Studies of this sort have been carried in the past
\cite{Bonning:2007vt, Komossa:2008qd, Bogdanovic:2008uz,
Heckman:2008en, Shields:2008kn, Strateva:2008wt, Decarli:2009av,
Lauer:2009us, Vivek:2009mm, Robinson:2010ui, Shields:2013yaa,
Decarli:2014wca} prompted by the numerical relativity simulations that
predicted large recoil velocities from the merger of binary black
holes \cite{Campanelli:2007ew, Campanelli:2007cga}.  A review of those
early efforts is summarized in \cite{Komossa:2012cy}.

This new case of the QSO 3C 186 is of particular interest since
its differential velocity (if interpreted in terms of gravitational
wave recoil) may be used to determine the parameters
of the progenitor binary black hole system and the final black
hole being ejected from the merged galaxies.

Full numerical simulations of the merger of binary black holes
have produced detailed predictions for the remnant final black
hole mass, spin and recoil velocity
\cite{Lousto:2009mf,Lousto:2012gt,Lousto:2013wta,Zlochower:2015wga}
and the probability of a given recoil velocity to be observed
\cite{Schnittman:2007sn,Lousto:2009ka,Lousto:2012su}.
Those ``phenomenological'' formulas relate the binary parameters of
the progenitor, i.e. individual masses and spins, to the final mass, spin and
(recoil) velocity of the merged hole with high accuracy. The large
value of the measured differential redshift of the
broad and narrow lines can only be the result
of the gravitational recoil if the progenitor binary had a 
mass ratio close to unity and highly spinning progenitor
black holes (BHs).
We can also determine the direction of the recoil velocity with respect
to the merger orbital plane.

These techniques, used here for QSO 3C 186, clearly apply to any
highly recoiling system, as for example those candidates cited above.
In order to cover different pre-merger
scenarios, we study binaries with parameters based on hot and cold
accretion models, as well as two gas-poor merger models, as shown in Fig.~\ref{fig:start}


\section{Results}\label{sec:Results}
\begin{figure}
  \includegraphics[width=.49\columnwidth]{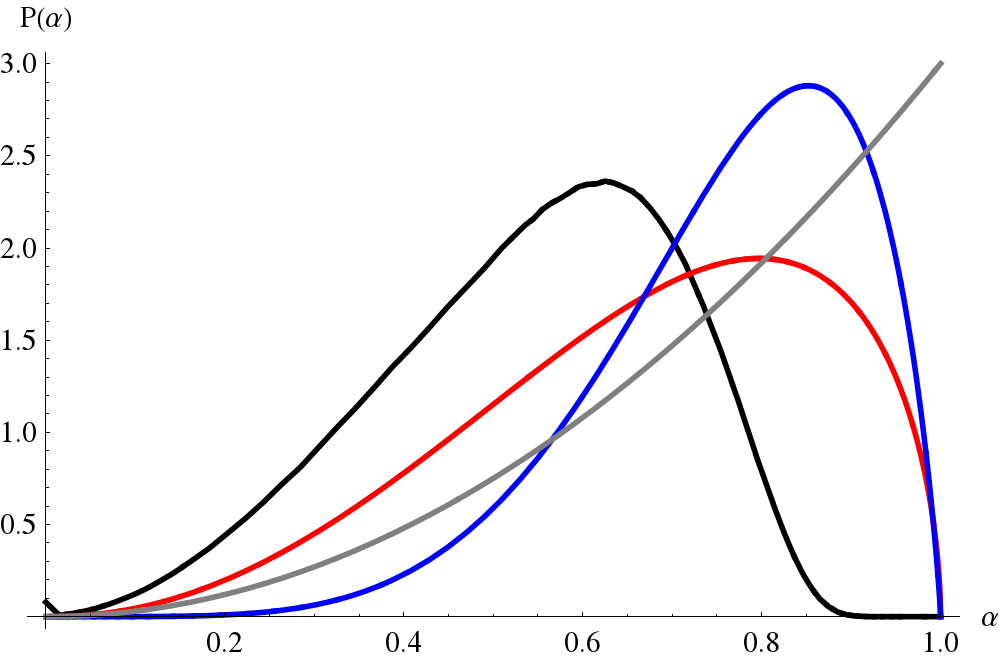}
  \includegraphics[width=.49\columnwidth]{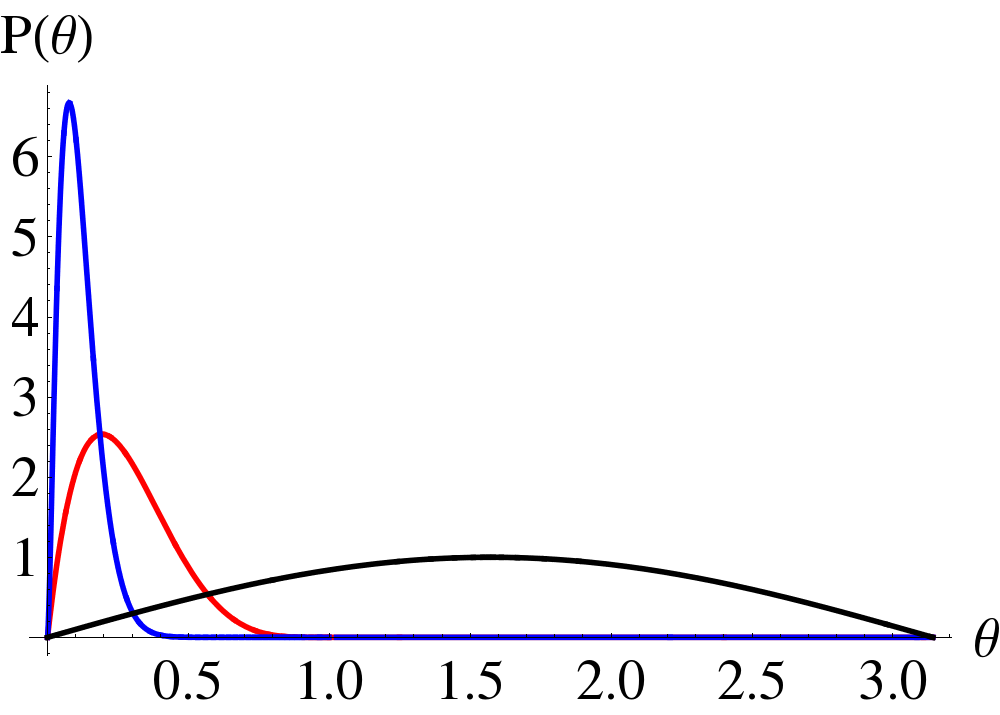}
  \caption{The distribution of spin magnitudes and polar orientations
    for {\it hot} (red), {\it cold} (blue), and {\it dry} (black) mergers based
    on~\cite{Lousto:2012su, Zlochower:2015wga}, as well as
  for the {\it uniform} (gray) distribution of spin magnitudes. Note that the
    both the {\it dry} and {\it uniform}  distributions are uniform in $\mu =
  \cos\theta$. The {\it hot} and {\it cold} distributions are strongly
peaked near $\theta=0$. The {\it uniform} distributions assumes equal
likelihood for any spin in the ball $|\vec \alpha|\leq1$. }\label{fig:start}
\end{figure}

For our statistical analysis, we consider the recoils from binaries sampled
from the following distributions. For the mass ratio, we use
a distribution motivated by
cosmological simulations,  $P(q) \propto q^{-0.3} (1-q)$,
as given in Ref.~\cite{Yu:2011vp, Stewart:2008ep, Hopkins:2009yy}. For
the spins, we consider four different distributions which we will
denote by {\it Hot}, {\it Cold}, {\it Dry}, and {\it Uniform}. The
{\it Hot} and {\it Cold} distributions are based on the hot and cold
accretion models given in
\cite{Lousto:2012su}. In these models, the merger is assumed to be gas
rich and the subsequent accretion both reorients the spins (towards
partial alignment) and induces relatively large spin magnitudes. The cold model, in particular, severely
constrains the polar orientations of the spin, which severely limits
the magnitude of the recoil. Because of the very low probability for
large recoils in the dry model, our sample size was $3.1\times10^{8}$
binaries for the cold model. For all other models, the sample size was $10^7$ binaries.
The dry model is based on
\cite{Zlochower:2015wga}. For the dry model, we assume accretion is
inefficient at aligning the spins, and thus assume a uniform
distribution of spin directions. The magnitude of the spin is
determined by assuming past mergers were also gas poor. This leads to
the spin-magnitude distribution shown in Fig.~\ref{fig:start}. Finally, the {\it Uniform} model simply
assumes uniform probabilities for the spin magnitudes and directions
within the unit sphere (i.e., uniform probabilities for all directions
and a probability density of $3 \alpha^2$ for the magnitude of the
spin). This uniform model has a strong bias towards high spins (that could be
the product of gas-rich pre-merger scenarios) combined with random
distributions of the spin directions (that could be the product
of anisotropic accretion,  see Figs. 4 and 7 in \cite{Perego:2009cw},
or retrograde circumbinary accretion \cite{Schnittman:2015eaa}). 
Thus, while the uniform distribution
is geometrical in origin, it represents a series of
astrophysically plausible scenarios and provides the most favorable
distributions for observing high recoils of thousand of km/s.

These distributions are summarized in Fig.~\ref{fig:start}.

To model the recoil, we use the formulas given
in~\cite{Zlochower:2015wga}. Based on those formulas, we can conclude
that a binary with mass ratio $q< 0.23$ {\em cannot} recoil as fast
as $2000\ \KMS$. This holds true regardless of the progenitors spin
magnitudes and orientations. If we further assume, for instance, the {\it dry}
distribution of spin magnitudes, then the mass ratio cannot be smaller
than $0.28$.  Thus, if the supermassive black hole
(SMBH) in QSO 3C 186 resulted from the {\it
quasicircular}
\footnote{Comparable masses binary black holes are very efficient
in reducing any initial eccentricity through radiation of gravitational
waves \cite{Peters:1964zz} down to the merger 
\cite{Mroue:2010re,Lousto:2015uwa}.}
 inspiral of two SMBHs, the progenitor BHs must have had
similar masses.

\begin{figure*}
  \includegraphics[width=.49\columnwidth]{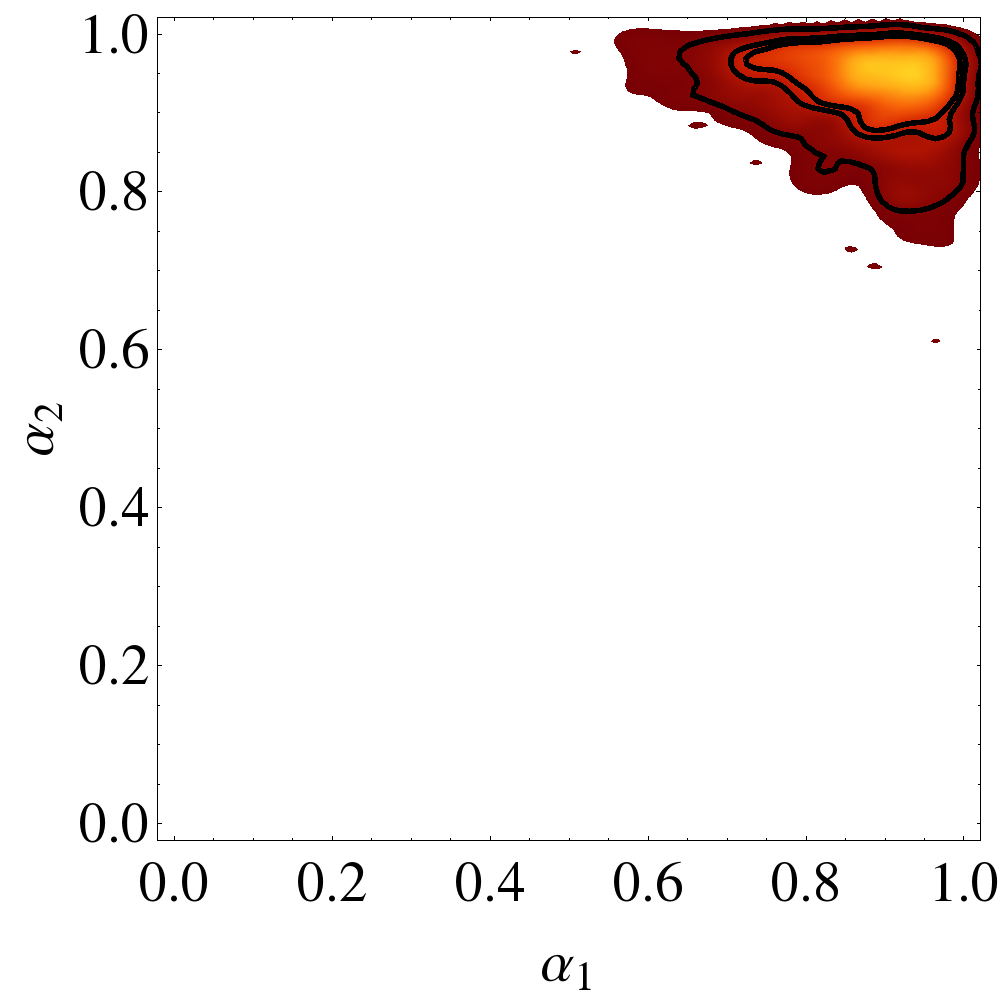}
  \includegraphics[width=.49\columnwidth]{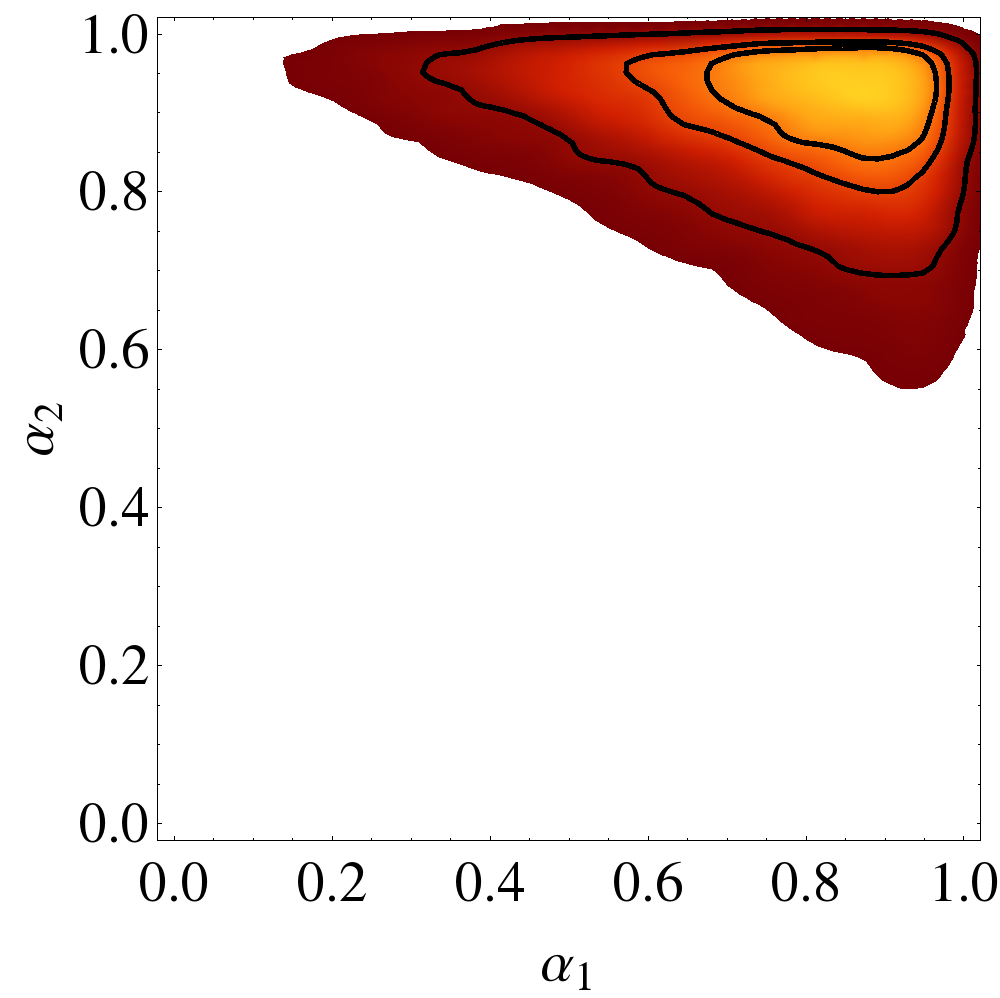}
  \includegraphics[width=.49\columnwidth]{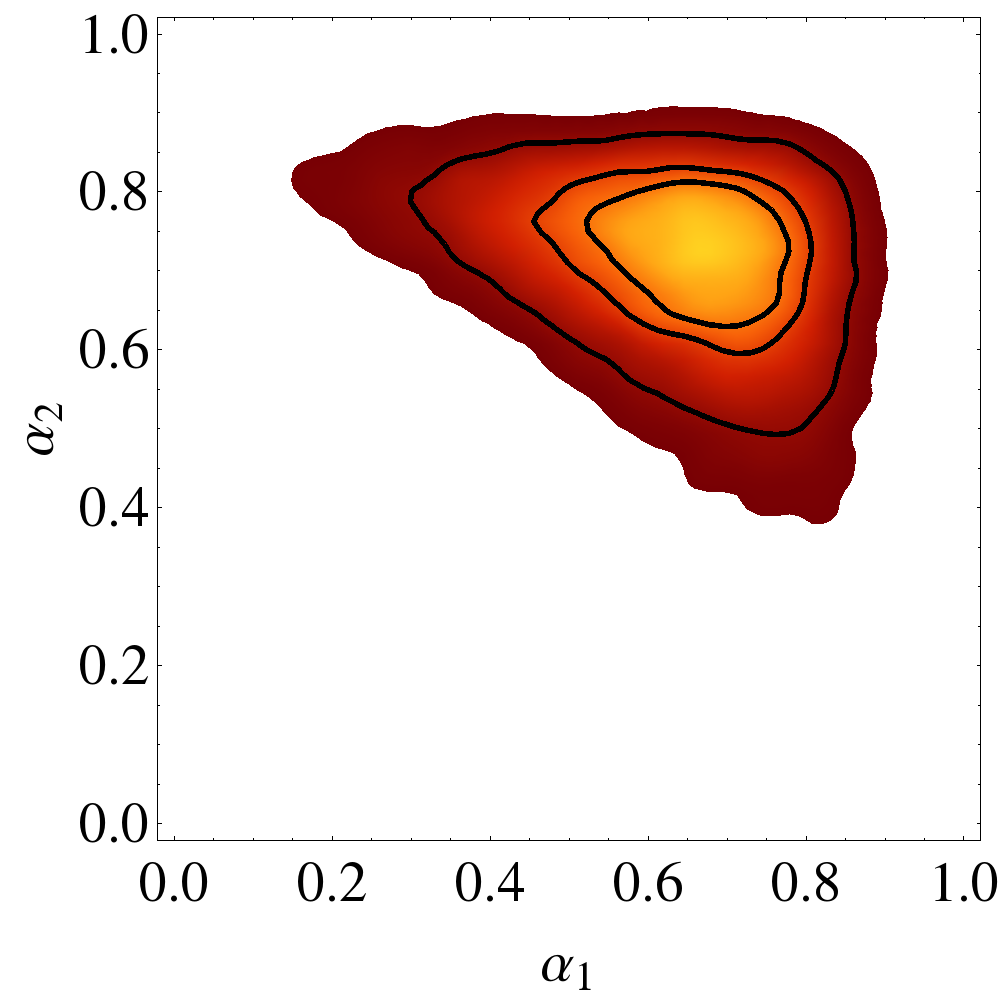}
  \includegraphics[width=.49\columnwidth]{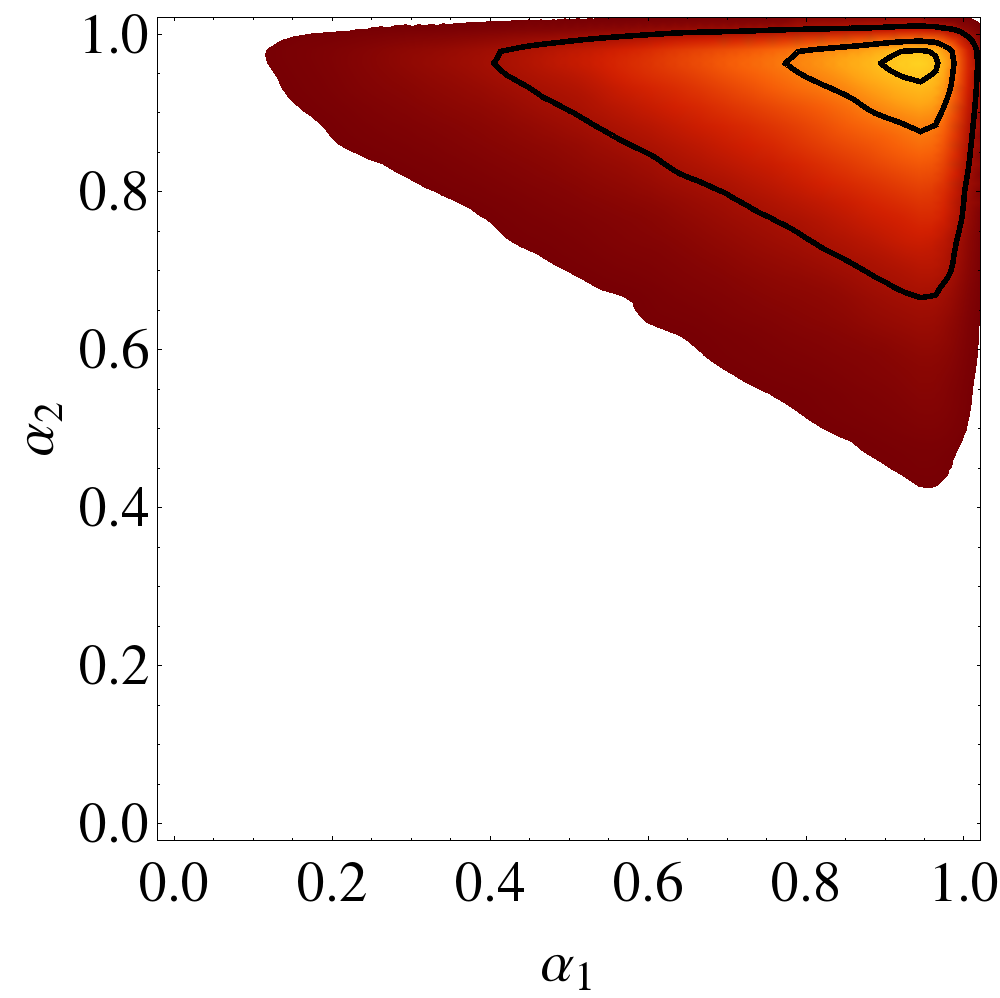}\\
  \\
  \includegraphics[width=.49\columnwidth]{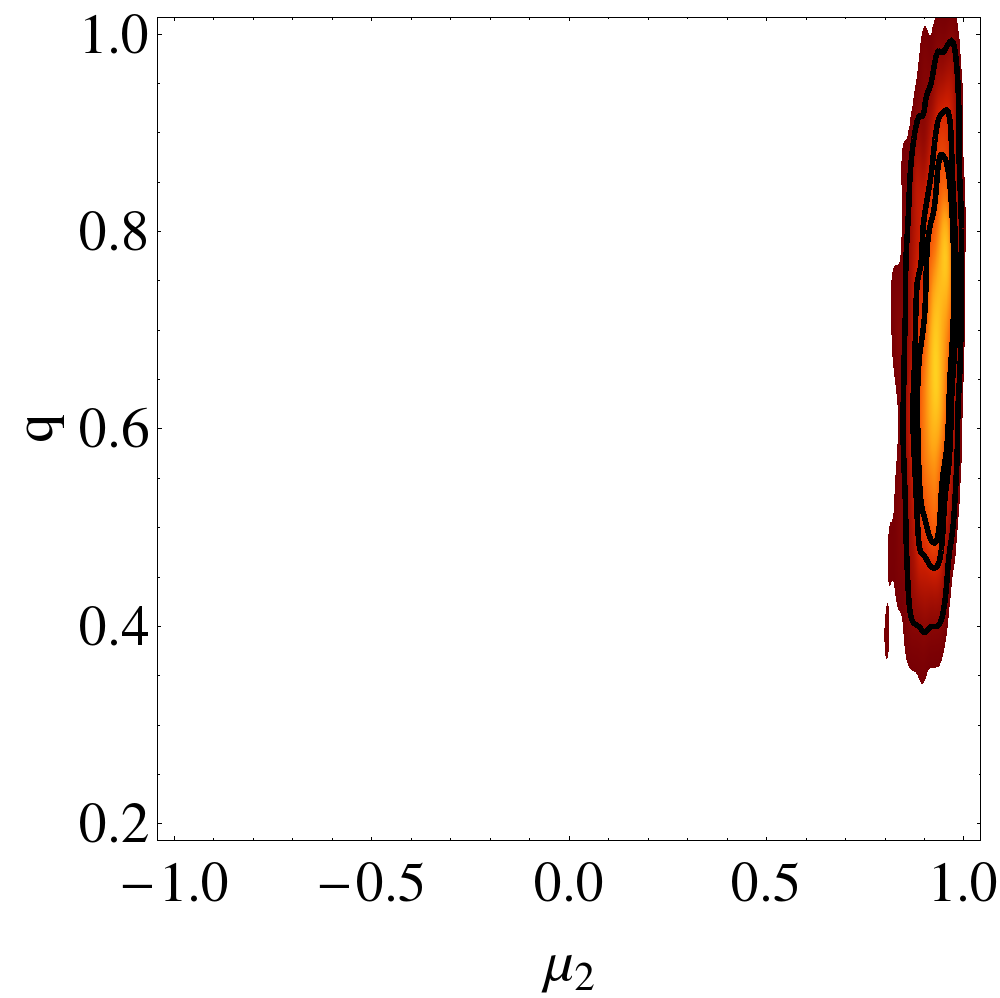}
  \includegraphics[width=.49\columnwidth]{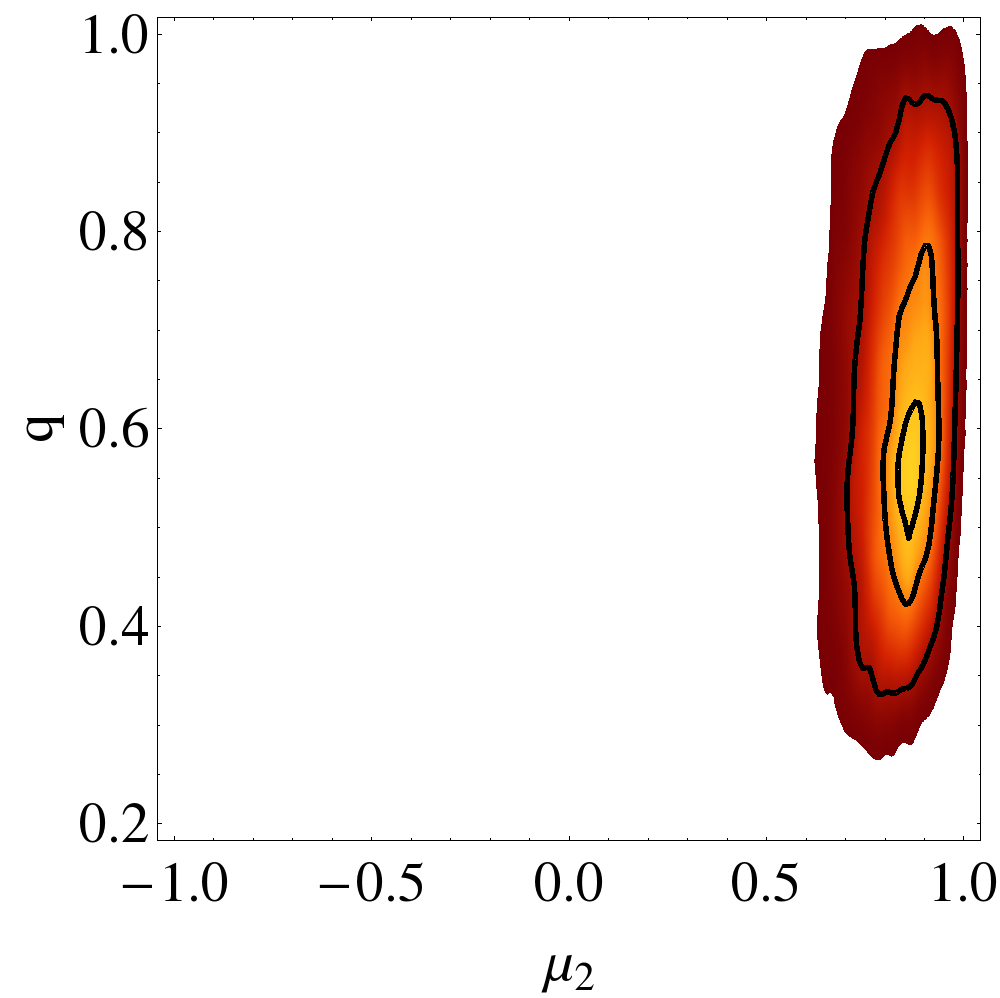}
  \includegraphics[width=.49\columnwidth]{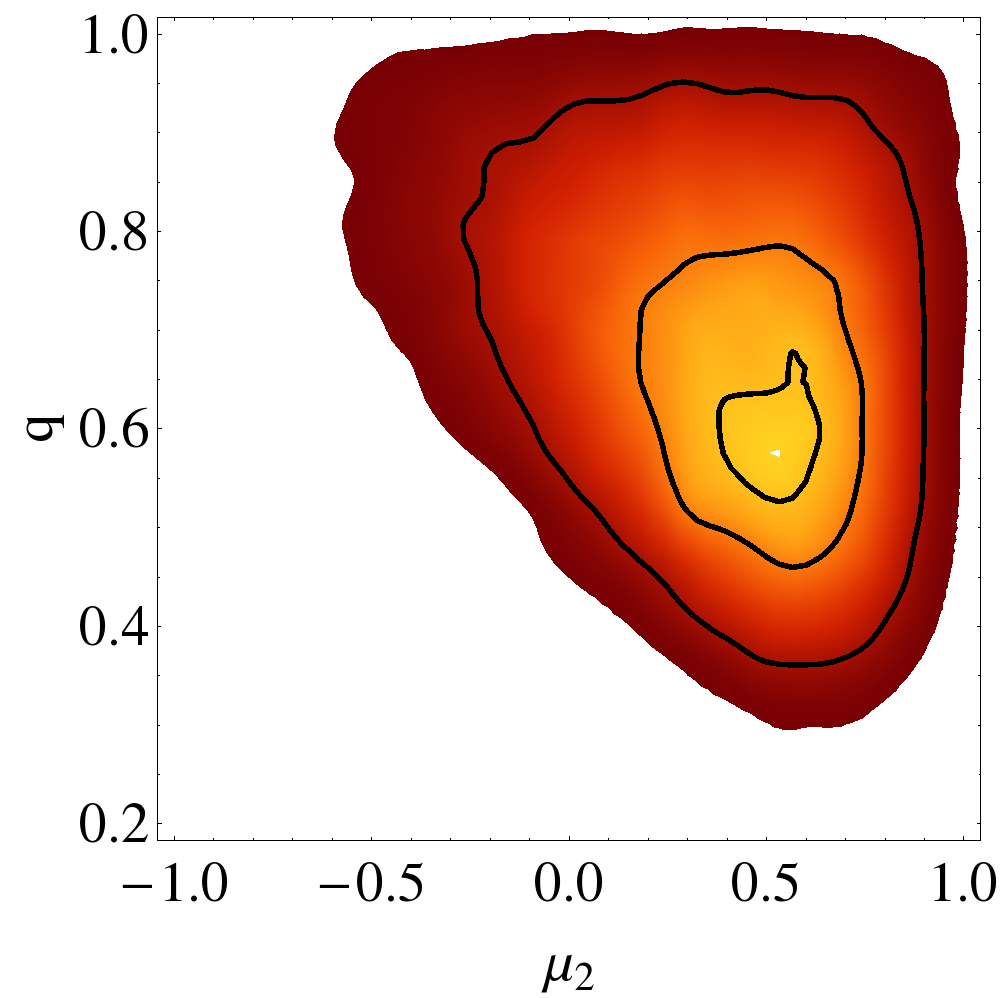}
  \includegraphics[width=.49\columnwidth]{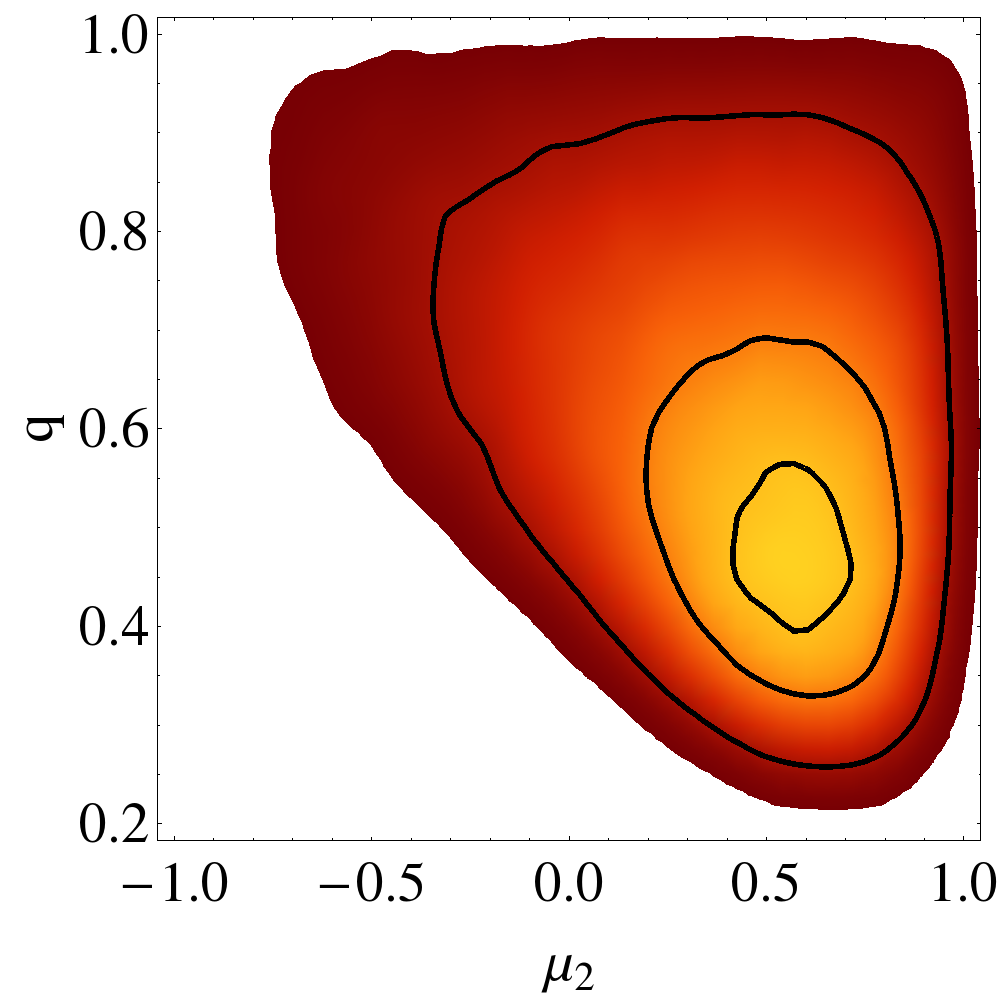}
  \\
  \includegraphics[width=.49\columnwidth]{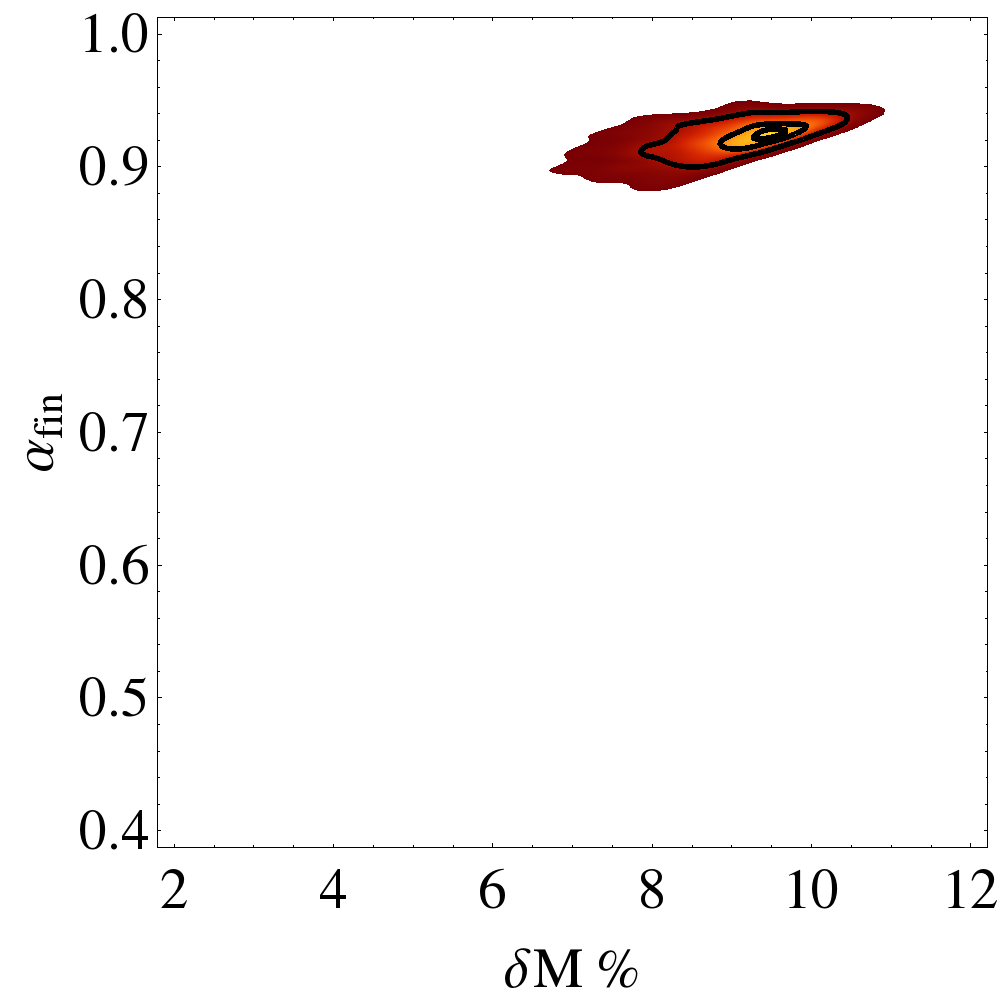}
  \includegraphics[width=.49\columnwidth]{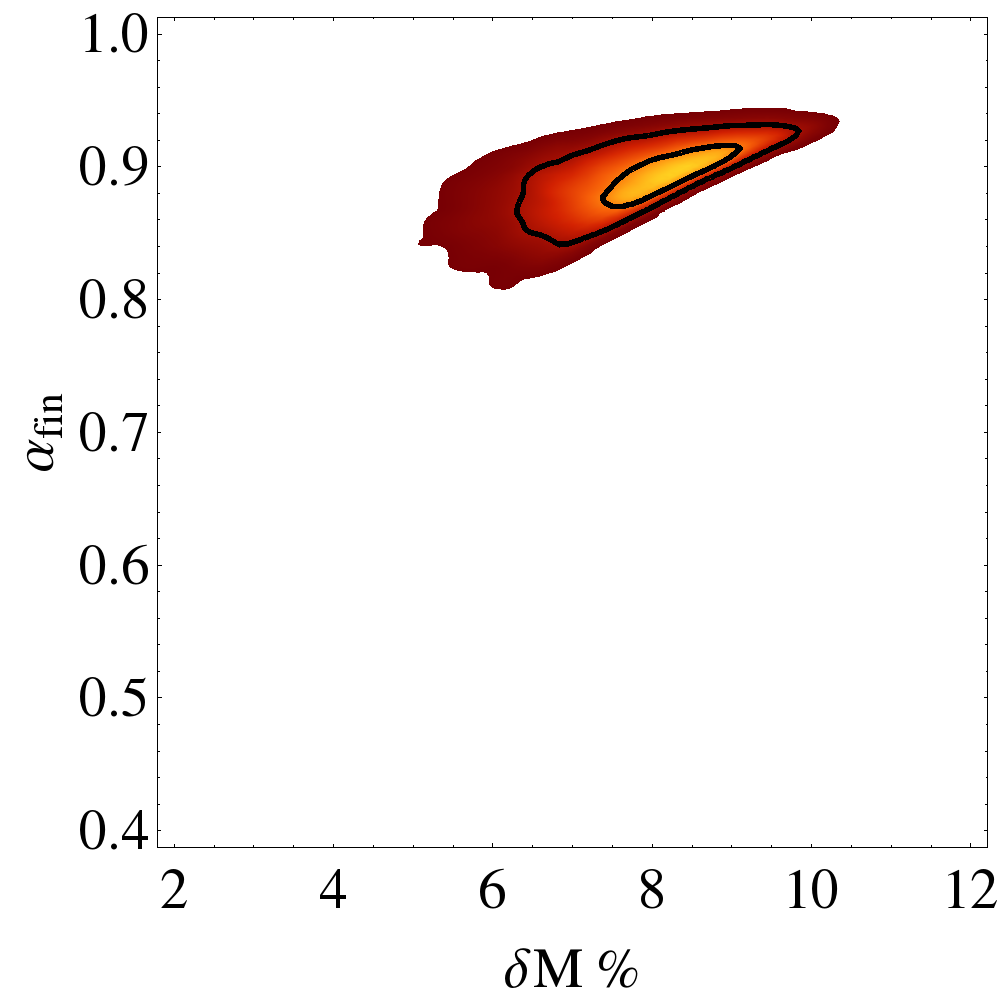}
  \includegraphics[width=.49\columnwidth]{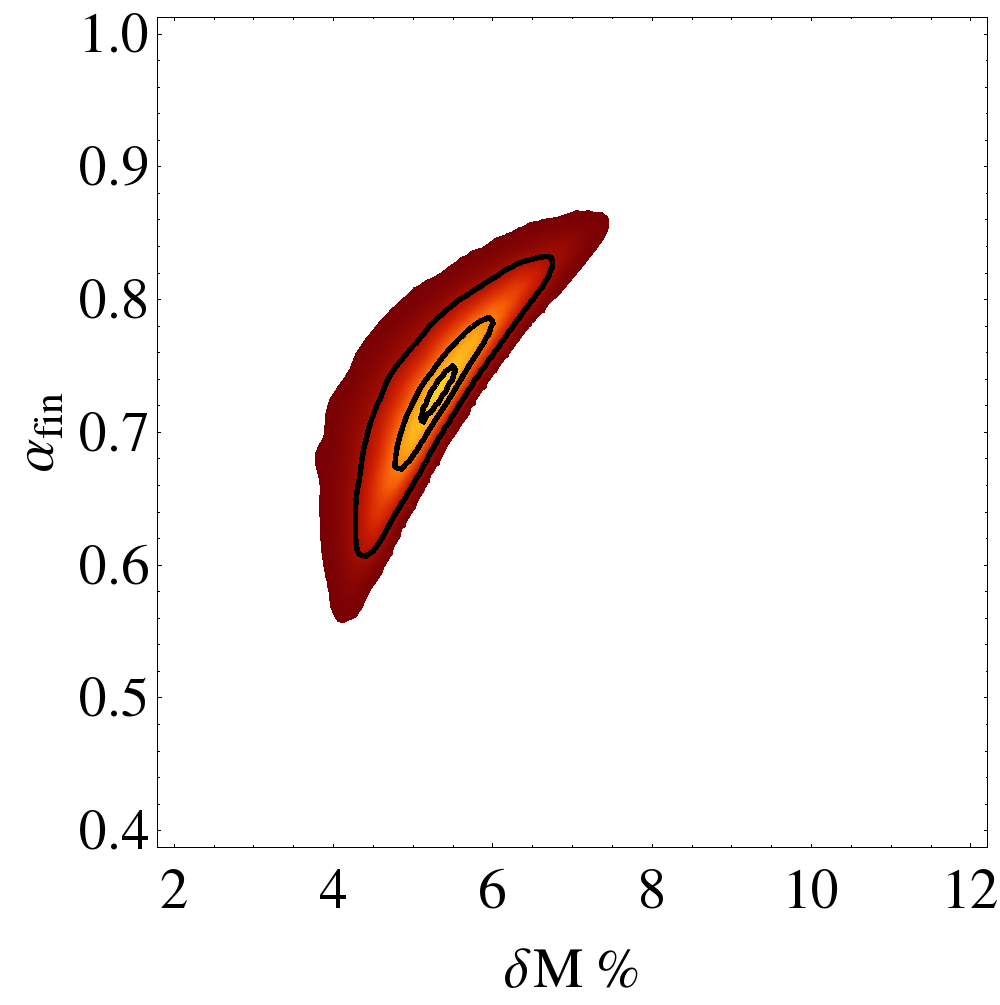}
  \includegraphics[width=.49\columnwidth]{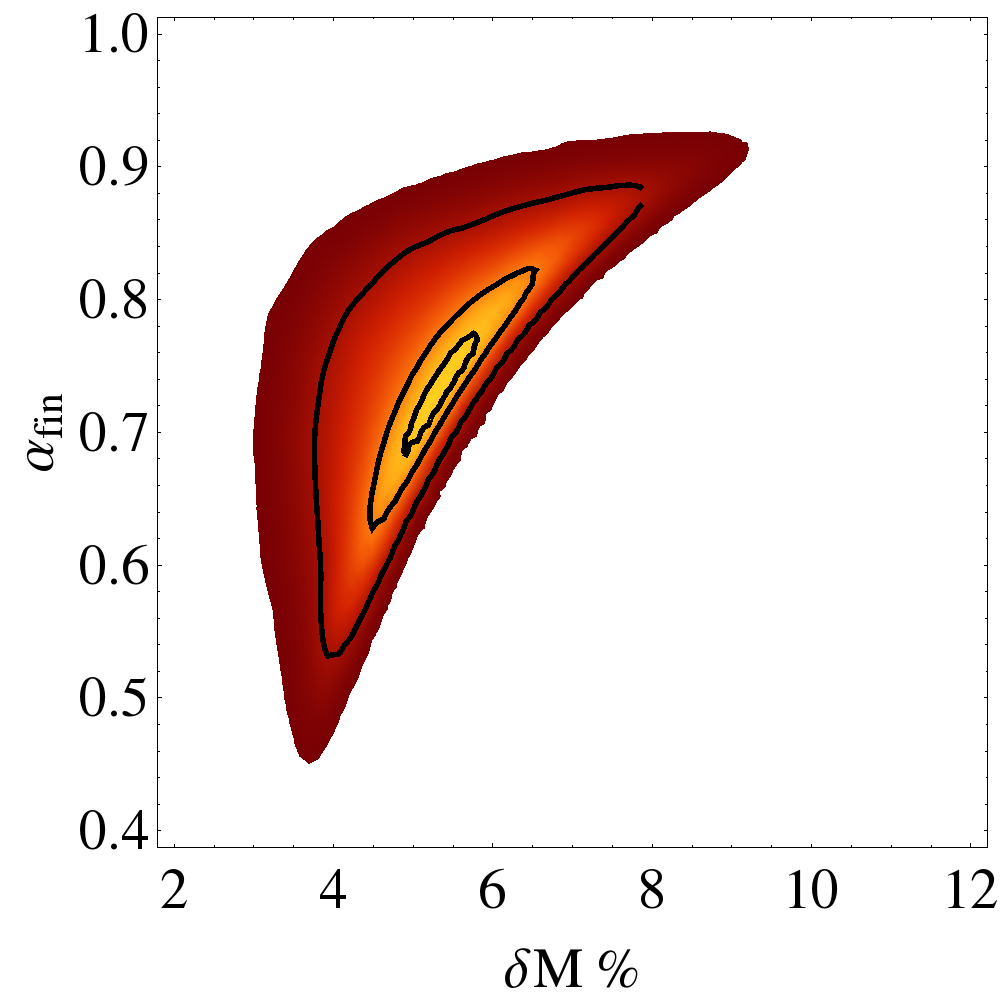}
 
  \caption{Relative probabilities for a recoil $2000\ \KMS$ or larger
   for configurations with a given spin magnitude $\alpha_1$ and $\alpha_2$ or polar
   orientation (of the larger BH) $\mu_2 = \cos \theta_2$ and mass ratio, $q$.
   The probabilities come from distributions (arranged left to right) due to cold accretion,
   hot accretion, uniform volume distributions based
   in dry merger spins, and a uniform
 distribution in spin magnitudes and directions.
 In each panel, there are contour lines at 0.9, 0.61, 0.14 of the
 maximum. The boundary of the shaded region is $0.01$ of the maximum.}
\label{fig:distributions}
\end{figure*}

Figure~\ref{fig:distributions} shows the probabilities for a recoil of
$2000\ \KMS$ or larger as a function of the two spins $\alpha_1$ and
$\alpha_2$ for each distribution, as well as the
probabilities as a function of mass ratio and the polar orientation of
the spin of the larger BH, $\mu_2=\cos\theta_2$. Note that the probability of any binary
recoiling at $2000\ \KMS$ or larger for the four models are
$3\times10^{-4}\%$, $0.19\%$, $0.23\%$, and $2.14\%$ for cold, hot, dry,
and uniform volume distributions, respectively. Finally, we show
probabilities for the remnant spin and total radiated mass (in terms
of the binary's initial mass).

Based on Fig.~\ref{fig:distributions}, we can estimate the parameters
of the progenitor binary. Assuming a dry, hot, cold, uniform volume merger,
we get the parameters in
Table~\ref{table:params}. Furthermore, using the empirical
formulas in~\cite{Zlochower:2015wga}, 
we find that the final remnant black hole spin and total radiated
energy. For instance, for the dry mergers, the final spin is
$\alpha_f = 0.75^{+0.08}_{-0.13}$
and the binary converted $5.5^{+1.1}_{-1.1}$\%
of its total mass into gravitational radiation.
The other models lead to even higher final remnant spins and radiated
gravitational energy.

\begin{table}
  \caption{Parameters of the progenitor binary assuming a dry, hot, cold, uniform volume
    merger. $\mu_{1,2} = \cos \theta_{1,2}$ is the cosine of the angle
    each spin makes with the direction of the orbital angular
    momentum, $\alpha_{1,2}$ are the dimensionless spin magnitudes, and
$q=m_1/m_2 \leq 1$ is the mass ratio. The errors are given at
1-$\sigma$ level.
We also provide the final spin of the merged hole and the total radiated
energy in units of the binary's total mass based on the
binary-parameter distributions in Fig.~\ref{fig:distributions}. Lastly, we
include the probabilities, given hot, cold Dry, or uniform
distributions for recoils of $2000\ \KMS$ or larger.
}\label{table:params}
\begin{ruledtabular}
  \begin{tabular}{llllllll}
    &   $\mu_1$ & $\mu_2$& $\alpha_1$& $\alpha_2$ & $q$\\
\hline\\
\vspace{-10pt}\\
\vspace{5pt}Cold & $0.99^{+0.01}_{-0.07}$ & $0.93^{+0.06}_{-0.07}$ & $0.94^{+0.06}_{-0.22}$ & $0.95^{+0.05}_{-0.09}$ & $0.70^{+0.29}_{-0.21}$\\
\vspace{5pt}
Hot & $1.00^{+0.00}_{-0.28}$ & $0.88^{+0.11}_{-0.19}$ & $0.78^{+0.21}_{-0.34}$ & $0.96^{+0.04}_{-0.24}$  & $0.60^{+0.36}_{-0.25}$\\
\vspace{5pt}Dry &  $0.52^{+0.42}_{-0.51}$ & $0.45^{+0.48}_{-0.44}$ & $0.74^{+0.08}_{-0.39}$ & $0.71^{+0.14}_{-0.14}$ & $0.56^{+0.39}_{-0.17}$ \\
%
\vspace{5pt}Uni. & $0.09^{+0.91}_{-0.08}$ & $0.63^{+0.31}_{-0.59}$ &
$1.00^{+0}_{-0.49}$ & $1.00^{+0}_{-0.29}$ & $0.55^{+0.28}_{-0.24}$\\
  \end{tabular}
  \vspace{5pt}
  \begin{tabular}{llll}
    &   $\alpha_{\rm final} $ & $\delta M\%$ & $P(v>2000\ \KMS)$ \\
\hline\\
\vspace{-10pt}{}\\
\vspace{5pt}Cold & $0.93^{+0.02}_{-0.03}$ & $9.6^{+0.8}_{-1.4}$ & $3\times10^{-4}$\%\\
\vspace{5pt}Hot & $0.90^{+0.03}_{-0.04}$& $8.6^{+1.3}_{-1.8}$ & 0.19\%\\
\vspace{5pt}Dry  & $ 0.75^{+0.08}_{-0.13}$ & $5.5^{+1.1}_{-1.1}$ & 0.23\%\\
%
\vspace{5pt}Uni. & $0.75^{+0.13}_{-0.19}$ & $5.5^{+2.3}_{-1.4}$ & 2.14\% 
\end{tabular}

\end{ruledtabular}
  \end{table}

  \section{Conclusion and discussion}\label{sec:Discussion}

While for the ideal configuration \cite{Lousto:2011kp} of equal-mass
binaries and maximally spinning black holes with spins at nearly
50-degrees from the orbital angular momentum and opposite phases,
recoil velocities can reach up to 5000km/s.  By demanding high
velocities, above 2000km/s, one can place important constraints on the
parameters of the progenitor binary.  We find that, independent
of the merger model we adopt, the mass ratio has to be $q>1/4$  and
that the spin of the holes are likely above $50\%$ of their maximum
value.  Note that according to Fig. 1, (bottom-right panel) of
\cite{Chiaberge:2016eqf} the presence of low signal to noise shells or
tidal tails in the host galaxy are typical of major galaxy merger
remnants, i.e. the two merging galaxies have masses that are equal to
within a factor of 3.

Those highly-recoiling configurations also require a misalignment of the spins with the
orbital angular momentum of the binary, which suggests that any
circumbinary accretion did not have enough time to align spins
\cite{Bogdanovic:2007hp,Miller:2013gya}, i.e. the merger was either
gas-poor or the black holes were too massive for accretion to make an
impact on the direction of the spins.
Note that the estimates of the final black hole mass in
\cite{Chiaberge:2016eqf} sets its value in the range
$3-6\times10^9M_{\odot}$.
Increasing the measured recoil velocity can dramatically narrow
the possible region of the binary parameter space, and
even exclude some of the pre-merger scenarios, such as the cold
accretion one.
This also suggest, that if one had an
independent way of measuring another parameter of the system, for
instance the final spin of the remnant, one could choose, based on
Table~\ref{table:params}or Fig.~\ref{fig:distributions} among the
different models for the pre-merger stage of the binary black hole.

If we assume that the 11kpc of offset between the AGN and the host
galaxy is due to a transverse component of the recoil velocity of
nearly 1000km/s, the time elapsed from the merger of the black hole
is around $10^7$ years.  This in turn, allows us to claim that our
bounds, based on recoil velocities $v>2000$km/s is a conservative one
(since the total recoil velocity, including transversal and potential
of the host galaxy components, would be even larger), and that the
actual parameters of the precursor binary are even closer to more
comparable masses and higher spins.

In relation to the above time scale another important factor to
consider is the lifetime of accretion disks carried by recoiling black
holes \cite{Blecha:2008mg,Blecha:2010dq}. In \cite{Chiaberge:2016eqf},
assuming a radiative efficiency of $\epsilon=0.1$ and the luminosity
and BH mass estimated for 3C 186, they derive a lifetime for the
disk of $t_{disk}\sim10^8$yr.  This is an order of magnitude longer
than the estimate above and hence the transverse velocities would not need
to be much larger than $100$ km/s for the accretion disk to survive 
until a 11kpc offset is reached.

Finally, large recoil velocities are strongly beamed along the orbital
angular momentum (see Figs. 11-14 of Ref.~\cite{Lousto:2012su} and
Fig. 7 in Ref.~\cite{Zlochower:2015wga}).  This means that we must be
seeing the system in a rather face-on angle with respect to the late
merger orbital plane. It is interesting to correlate this with the
radio, optical and x-ray maps of QSO 3C 186.

\acknowledgments The authors thank T.Bogdanovic, E.Bonning, M.Chaberge, Cole-Miller, M.Dotti, J.Krolik and
J.Schnittman for discussions and gratefully acknowledge the NSF for
financial support from NSF Grants No.  PHY-1607520, No. ACI-1550436,
No. AST-1516150, and No. ACI-1516125.

\end{document}